\RequirePackage{fix-cm}
\documentclass[twocolumn,natbib,epjc3]{svjour3}
\smartqed
\usepackage{graphicx}
\usepackage{amsmath,amssymb,bm}
\usepackage{array}
\usepackage{float}
\usepackage{mathptmx}
\usepackage[T1]{fontenc}
\usepackage[utf8]{inputenc}
\setcitestyle{numbers,square,comma,sort&compress}
\usepackage[hidelinks]{hyperref}
\hypersetup{hypertexnames=false,pdftitle={Gravitational-Wave Propagation in Quadratic Einstein--Cartan--Dirac Gravity: A Conditional Tensor-Sector Analysis}}

\journalname{Eur. Phys. J. C}

\begin{document}

\title{Gravitational-Wave Propagation in Quadratic Einstein--Cartan--Dirac Gravity: A Conditional Tensor-Sector Analysis}
\titlerunning{Gravitational-Wave Propagation in EC--Dirac Gravity}
\authorrunning{Zhi-Fu Gao et al.}

\author{Zhi-Fu Gao\thanksref{addr1}
\and Biao-Peng Li\thanksref{addr1}
\and Na Wang\thanksref{addr1}
\and Hui Wang\thanksref{e1,addr2}
\and Guo-Qiang Jin\thanksref{addr3}
\and Zhoujian Cao\thanksref{addr4,addr5}
\and Luiz C. Garcia de Andrade\thanksref{addr6,addr7}}
\thankstext{e1}{Corresponding author. E-mail: wanghuiycu@163.com}

\institute{State Key Laboratory of Radio Astronomy and Technology, Xinjiang Astronomical Observatory, CAS, 150 Science, 1-Street, Urumqi, Xinjiang 830011, China\hypertarget{addr1}{}\label{addr1}
\and Shanxi Province Intelligent Optoelectronic Sensing Application Technology Innovation Center, Yuncheng University, Yuncheng 044000, China\hypertarget{addr2}{}\label{addr2}
\and College of Mechanical and Electrical Engineering, Tarim University, Alar, Xinjiang 843300, China\hypertarget{addr3}{}\label{addr3}
\and Institute for Frontiers in Astronomy and Astrophysics, Beijing Normal University, Beijing 102206, China\hypertarget{addr4}{}\label{addr4}
\and School of Physics and Astronomy, Beijing Normal University, Beijing 100875, China\hypertarget{addr5}{}\label{addr5}
\and Departamento de Física Teórica, IF-UERJ, Rua São Francisco Xavier 524, Maracanã, Rio de Janeiro, RJ 20550, Brazil\hypertarget{addr6}{}\label{addr6}
\and Institute of Cosmology and Philosophy of Nature, Krizževci 48260, Croatia\hypertarget{addr7}{}\label{addr7}}


\date{}
\maketitle

\begin{abstract}
We study leading tensor propagation in a specified first-order Einstein--Cartan--Dirac theory with the parity-even quadratic torsion invariant $\zeta T_{\lambda\mu\nu}T^{\lambda\mu\nu}$. Eliminating the nondynamical Lorentz connection on the regular algebraic branch gives a metric description containing the Einstein--Hilbert term, the Dirac action, and a $\zeta$-dependent axial--axial fermion contact interaction. We evaluate the transverse-traceless (TT) Hessian under an explicit factorized mean-field closure, with vanishing connected-current TT response, vanishing first- and second-order TT response of the factorized current scalar, and zero intrinsic TT anisotropic stress. Under these imposed conditions, the leading two-derivative tensor action has $Q_T=M_{\rm Pl}^{2}$ and $c_T=1$, with no leading dispersion or propagation-induced polarization splitting. These conditions do not follow from isotropy of the one-point axial current, and the present calculation does not derive them from a microscopic fermionic state. The result is therefore conditional and does not establish a general Einstein--Cartan--Dirac no-go theorem. We identify the connected-current TT response that must be calculated to extend the result beyond this closure. Published NANOGrav summaries are used only to check consistency with the GR-like prediction; no new timing-data likelihood or constraint on $\zeta$ is obtained.
\end{abstract}

\section{Introduction}
\label{sec:intro}

General relativity (GR) has passed a wide range of precision tests, yet extensions involving spacetime torsion remain well motivated because they couple geometry directly to the intrinsic spin of matter \citep{Kibble1961,Hehl1976,Shapiro2002,Trautman1973}. In Einstein--Cartan (EC) gravity, the tetrad and Lorentz connection are independent variables and, in the minimal two-derivative theory, torsion is constrained algebraically rather than propagating as an additional wave degree of freedom. Minimally coupled Dirac matter sources the axial irreducible component of torsion, and eliminating the connection generates a local axial--axial fermion interaction \citep{Kibble1961,Hammond2002}.

Quadratic torsion invariants provide a natural extension of this setting because they change the algebraic response of the connection to the spin current. Spin--torsion cosmologies have also been studied in homogeneous Einstein--Cartan and spin-fluid settings, where the matter-sector realization and its dilution history play an essential role \citep{Tsamparlis1979,Poplawski2012,Kranas2019}. This is conceptually different from theories containing derivative torsion, curvature-squared terms, or independent propagating connection modes. The distinction matters directly for gravitational-wave (GW) phenomenology: a nonzero torsion background, or a torsion-dependent contribution to the matter stress tensor, does not by itself imply a modified tensor characteristic, a graviton mass, or a frequency-dependent propagation law. The physical tensor operator must instead be derived after the auxiliary connection has been consistently eliminated.

Previous work illustrates several complementary possibilities. Gravitational-wave generation in Einstein--Cartan theory has been studied in the post-Newtonian regime \citep{Battista2021}, while weak-torsion propagation analyses in Riemann--Cartan backgrounds can retain a null leading characteristic and modify subleading amplitude or polarization transport \citep{Elizalde2023}. Cosmological scalar and tensor perturbations in Einstein--Cartan settings have also recently been revisited from a perturbation-theory perspective \citep{Ranjbar2024}. The present calculation addresses a different question: for a Dirac-sourced \emph{auxiliary} torsion field with the specific parity-even quadratic invariant in Eq.~\eqref{eq:grav_action}, which terms survive in the physical TT Hessian after the Cartan constraint is imposed? This action-level reduction is what permits an unambiguous match to PTA propagation observables. A related ordering of the constrained variation has been used in dynamical Barbero--Immirzi gravity \citep{Gao2026BI}, but the field content, the operator controlling tensor normalization, and the standard-siren observable in that setting are distinct from the Dirac-sourced algebraic sector and PTA kernel studied here. Parity-violating theories can distinguish the two helicities and generate birefringent propagation \citep{Alexander2009,Nishizawa2018}; more general modified-propagation frameworks can also alter phase, amplitude, or lensing observables \citep{Ezquiaga2020prop}. These mechanisms are observationally distinct and should not be inferred solely from the presence of torsion.

The scope of this work is specific. The algebraic elimination of nondynamical torsion and the resulting matter contact interactions are established features of Einstein--Cartan theories. A broad class of models with nondynamical torsion has also been shown to contain only the massless graviton in the gravitational spectrum \citep{PhysRevD.104.064036}. We do not claim these general results as new.

For the particular invariant $T_{\lambda\mu\nu}T^{\lambda\mu\nu}$, we derive the $\zeta$-dependent axial--axial contact interaction and evaluate the leading constrained TT Hessian after imposing the algebraic connection equation. The tensor result is conditional on the factorized mean-field closure and the specified vanishing matter responses. We make the omitted connected-current TT response explicit; its value requires a microscopic fermionic state and is not calculated here. The contribution is therefore a model-specific constrained-Hessian calculation with an explicit statement of its matter-response assumptions.

The PTA discussion translates the conditional tensor prediction into observables and compares it with published summaries. Since the predicted leading propagation is GR-like, agreement with those summaries is a consistency check shared with GR. It does not identify algebraic torsion or constrain $\zeta$. A robust non-GR propagation signal could test or disfavor the minimal branch, subject to the assumptions stated above. Sections~\ref{sec:bg} and \ref{sec:tensor} derive the reduced action and tensor operator; Sec.~\ref{sec:numerical} presents the PTA consistency checks; Sec.~\ref{sec:discussion} discusses the scope and limitations; and Sec.~\ref{sec:conclusion} summarizes the results.
\section{Dirac-Induced Torsion in an FRW Universe}
\label{sec:bg}

We consider a spatially flat FRW background with an independent Lorentz connection and Dirac-sourced algebraic torsion. Because the connection has no derivative kinetic term, it must be eliminated by its field equation before the physical TT operator is identified.

\subsection{Action and Algebraic Connection Reduction}
\label{subsec:action}

We use the metric signature $(-,+,+,+)$ and define
\begin{equation}
T^{\lambda}{}_{\mu\nu}
\equiv
\Gamma^{\lambda}{}_{\mu\nu}-\Gamma^{\lambda}{}_{\nu\mu},
\qquad
\kappa\equiv 8\pi G=M_{\rm Pl}^{-2}.
\label{eq:torsion_def}
\end{equation}
The first-order Einstein--Cartan--Dirac action considered in this work is
\begin{equation}
\begin{aligned}
S &= \int d^{4}x\,e\Bigg[
\frac{1}{2\kappa}\Bigl(R(\omega)\\
&\qquad+\zeta T_{\lambda\mu\nu}T^{\lambda\mu\nu}\Bigr)
+\mathcal{L}_{D}(e,\omega,\psi)\Bigg],
\end{aligned}
\label{eq:grav_action}
\end{equation}
where $e=\det(e^{a}{}_{\mu})$, $\omega^{ab}{}_{\mu}$ is the independent Lorentz connection, and $\zeta$ is a dimensionless quadratic-torsion coupling. The minimally coupled Hermitian Dirac Lagrangian is
\begin{equation}
\mathcal{L}_{D}
=
\frac{i}{2}
\left[
\bar\psi\gamma^{a}e_{a}^{\mu}D_{\mu}\psi
-
(D_{\mu}\bar\psi)\gamma^{a}e_{a}^{\mu}\psi
\right]
-m\bar\psi\psi .
\label{eq:dirac_action}
\end{equation}
No additional axial coupling is appended to Eq.~\eqref{eq:dirac_action}; the coupling of the Dirac axial current to torsion is already contained in the full spin connection entering $D_{\mu}$.

Writing the connection as the Levi--Civita part plus contortion and decomposing torsion into its irreducible trace, axial, and tensor pieces, the minimally coupled Dirac field sources only the axial component \citep{Hehl1976,Shapiro2002}. It is convenient to define the axial torsion vector
\begin{equation}
A^{\mu}
\equiv
-\frac{1}{6}\epsilon^{\mu\nu\rho\sigma}T_{\nu\rho\sigma},
\qquad
J_{5}^{\mu}
\equiv
\bar\psi\gamma^{\mu}\gamma^{5}\psi .
\label{eq:axial_defs}
\end{equation}
We use the orientation $\epsilon^{0123}=+1/e$ (hence $\epsilon_{0123}=-e$ for the $(-,+,+,+)$ signature), $\{\gamma^{a},\gamma^{b}\}=2\eta^{ab}$, $\gamma^{5}=i\gamma^{0}\gamma^{1}\gamma^{2}\gamma^{3}$, and $\gamma^{ab}\equiv\tfrac12[\gamma^{a},\gamma^{b}]$. The spinor covariant derivatives are
\begin{equation}
\begin{aligned}
D_{\mu}\psi
&=\left(\partial_{\mu}+\frac14\omega_{ab\mu}\gamma^{ab}\right)\psi,\\
D_{\mu}\bar\psi
&=\partial_{\mu}\bar\psi-\frac14\omega_{ab\mu}\bar\psi\gamma^{ab},
\end{aligned}
\label{eq:spinor_conventions}
\end{equation}
which fix the relative sign of the axial torsion coupling below.
The algebraic connection sector is most transparent in irreducible torsion variables. Define the trace vector $V_\mu\equiv T^{\lambda}{}_{\mu\lambda}$ and a tensor $Q_{\lambda\mu\nu}$ satisfying $Q^{\lambda}{}_{\mu\lambda}=0$ and $\epsilon^{\mu\nu\rho\sigma}Q_{\nu\rho\sigma}=0$. Up to the total divergence $-2\nabla_\mu V^\mu$, its quadratic form is
\begin{equation}
\begin{aligned}
\mathcal L_{\rm conn}^{(2)}
&=\frac{1}{2\kappa}\Biggl[
\frac{2}{3}(\zeta-2)V_\mu V^\mu\\
&\qquad+\frac{3}{2}(1-4\zeta)A_\mu A^\mu\\
&\qquad+\frac{1}{2}(1+2\zeta)Q_{\lambda\mu\nu}Q^{\lambda\mu\nu}
\Biggr].
\end{aligned}
\label{eq:main_torsion_form}
\end{equation}
The zeros at $\zeta=2$, $1/4$, and $-1/2$ mark loss of algebraic invertibility in the trace, sourced axial, and tensor sectors, respectively; none creates a propagating torsion mode in this action. Appendix~A gives the convention-level decomposition.
With the conventions of Eqs.~\eqref{eq:torsion_def}--\eqref{eq:axial_defs}, the part of the Lagrangian that contains the sourced axial torsion can be written as
\begin{equation}
\mathcal{L}_{A}
=
\frac{3(1-4\zeta)}{4\kappa}
A_{\mu}A^{\mu}
+
\frac{3}{4}A_{\mu}J_{5}^{\mu}.
\label{eq:axial_lagrangian}
\end{equation}
The coefficient in Eq.~\eqref{eq:axial_lagrangian} follows directly from the Lorentz-irreducible decomposition of the torsion terms; the convention check, including the trace and tensor sectors, is given in Appendix~A. The axial connection equation is therefore algebraic,
\begin{equation}
A_{\mu}
=
-\frac{\kappa}{2(1-4\zeta)}J_{5\mu},
\qquad
\zeta\neq\frac14.
\label{eq:axial_solution}
\end{equation}
At $\zeta=1/4$ the axial quadratic coefficient vanishes, so the sourced connection equation no longer determines $A_{\mu}$; for a nonzero Dirac axial current it instead degenerates into a constraint on the source. The reduced solution in Eq.~\eqref{eq:axial_solution} therefore does not extend through this point. Moreover, as $|1-4\zeta|$ becomes small, the coefficient of the reduced four-fermion interaction becomes parametrically large, so the vicinity of the degeneracy is not interpreted here as a controlled weak-coupling mean-field regime. The Dirac-sourced axial reduction requires $\zeta\neq1/4$. The complete algebraic connection is regular for $\zeta\notin\{2,1/4,-1/2\}$; the unsourced trace and tensor sectors at $\zeta=2$ and $-1/2$ are algebraic flat directions, not additional propagating modes. The benchmark used below lies well inside this regular branch.

Substituting Eq.~\eqref{eq:axial_solution} back into the first-order action gives the equivalent metric description
\begin{equation}
\begin{aligned}
S_{\rm eff}
&=\int d^{4}x\sqrt{-g}\Biggl[
\frac{R(g)}{2\kappa}+\mathcal{L}_{D}(g,\psi)\\
&\qquad-\frac{3\kappa}{16(1-4\zeta)}J_{5\mu}J_{5}^{\mu}\Biggr],
\end{aligned}
\label{eq:reduced_action}
\end{equation}
for the regular axial branch. Thus the quadratic $T^{2}$ operator changes the coefficient of the local axial--axial interaction, but it does not introduce a derivative kinetic term for torsion or an additional propagating connection degree of freedom. This distinction is essential below: a background-dependent contact interaction can affect the matter sector and the cosmological background without generating, by itself, a new tensor mass, tensor sound speed, or frequency-dependent gravitational-wave dispersion.

\subsection{Homogeneous and Isotropic Ensemble}
\label{subsec:homogeneous}

For a spatially flat FRW background, we write the coarse-grained axial-current expectation value as
\begin{equation}
\left\langle J_{5}^{\mu}\right\rangle
=
\bigl(\rho_{5}(t),0,0,0\bigr).
\label{eq:axial_current}
\end{equation}
Here $\rho_{5}$ denotes the coarse-grained timelike axial-current density of the ensemble; Eq.~\eqref{eq:axial_current} is not identified with the spin four-vector of a single regular classical Dirac particle. Since the expectation value has no spatial component, it selects no preferred spatial direction and is compatible with the rotational symmetry of the FRW background. Its nonzero time component can nevertheless encode a macroscopic chiral or parity asymmetry of the underlying fermionic state; spatial isotropy therefore does not require parity symmetry of the ensemble.

On this branch Eq.~\eqref{eq:axial_solution} fixes the coarse-grained axial torsion algebraically in terms of $\rho_{5}$. The reduced interaction depends on the composite operator $J_{5\mu}J_{5}^{\mu}$. The one-point quantity $\bar J_{5}^{\mu}\equiv\langle J_{5}^{\mu}\rangle$ and the composite expectation value entering the reduced interaction are logically distinct: in a quantum or statistical fermionic state one need not have $\langle J_{5\mu}J_{5}^{\mu}\rangle=\bar J_{5\mu}\bar J_{5}^{\mu}$. Define the contracted connected contribution by $\mathcal C_{5}\equiv\langle J_{5\mu}J_{5}^{\mu}\rangle-\bar J_{5\mu}\bar J_{5}^{\mu}$. For Proposition~1 we explicitly adopt a factorized mean-field closure, setting $\mathcal C_{5}=0$ and neglecting its TT response; we additionally require the factorized scalar $\mathcal Q_{5,\rm MF}\equiv\bar J_{5\mu}\bar J_{5}^{\mu}$ to have zero first- and second-order TT response. These are extra state-level assumptions, not consequences of isotropy of the one-point current. A generic connected correlator with TT response can contribute to the tensor Hessian and is not evaluated here. A self-consistent cosmological stress tensor requires the microscopic state and its correlations to be specified.
For clarity, the omitted state-dependent term can be displayed explicitly. Let $S_{AA}$ denote the last term in Eq.~\eqref{eq:reduced_action}, and define $C_{AA}(\zeta)\equiv-3\kappa/[16(1-4\zeta)]$. Along the tensor-only synchronous path, $\sqrt{-g}=a^{3}$ is TT-invariant. If the factorized scalar has zero TT response, a generic connected contribution therefore enters as
\begin{equation}
\left.\delta_{\rm TT}^{\,n}S_{AA}\right|_{\rm conn}
=C_{AA}(\zeta)\int d^{4}x\,a^{3}\,\delta_{\rm TT}^{\,n}\mathcal C_{5},
\qquad n=1,2.
\label{eq:connected_tt_response}
\end{equation}
The proposition sets these connected responses to zero through its closure; Eq.~\eqref{eq:connected_tt_response} identifies the additional state-level quantity needed for a more general result.

The key propagation question is therefore not whether the background torsion is nonzero, but whether the reduced action in Eq.~\eqref{eq:reduced_action} contains a new transverse-traceless kinetic, gradient, mass, or parity-odd operator. We answer this question directly at quadratic order in the next section.

\section{Tensor Perturbations and Propagation}
\label{sec:tensor}

\subsection{Algebraic Torsion and the Tensor Quadratic Action}
\label{sec:tensor_action}

Because the Lorentz connection is auxiliary, the tensor Hessian is evaluated on the constrained branch. On the regular branch the connection Hessian is invertible, so eliminating $\omega$ before perturbing Eq.~\eqref{eq:reduced_action} gives the Schur complement of the full tetrad--connection Hessian. With $\Phi=(e,\psi)$,
\begin{equation}
S_{{\rm eff},\Phi\Phi}
=
S_{\Phi\Phi}
-
S_{\Phi\omega}S_{\omega\omega}^{-1}S_{\omega\Phi},
\qquad
\left.\frac{\delta S}{\delta\omega}\right|_{\omega_\star(\Phi)}=0,
\label{eq:schur_hessian}
\end{equation}
with all blocks evaluated at $\omega=\omega_\star(\Phi)$. This eliminates only the auxiliary connection, not fermionic perturbations. The TT projection below additionally imposes the factorized composite-current closure, zero first- and second-order TT response of $\mathcal Q_{5,\rm MF}$, and no intrinsic TT anisotropic stress; none follows from one-point isotropy alone.

\paragraph{Proposition 1 (leading constrained TT propagation).}
On the fully regular algebraic branch $\zeta\notin\{-1/2,1/4,2\}$, assume a spatially flat homogeneous and isotropic background with $\bar J_5^\mu=\rho_5u^\mu$, and adopt the factorized closure $\mathcal Q_{5,\rm MF}=\bar J_{5\mu}\bar J_5^\mu$ for the contact operator, neglecting the connected contribution $\mathcal C_5$. Also impose $\delta_{\rm TT}\mathcal Q_{5,\rm MF}=\delta_{\rm TT}^{2}\mathcal Q_{5,\rm MF}=0$ and no intrinsic TT anisotropic stress. A sufficient kinematic realization of the zero-response condition on this tensor-only path is $\bar J_5^\mu=\rho_5\delta^\mu_0$ with $\delta_{\rm TT}\rho_5=\delta_{\rm TT}^{2}\rho_5=0$; this is an imposed perturbative closure, not a microscopic derivation of a fermionic ensemble. Under these explicit closure and response assumptions, the leading two-derivative TT Hessian of Eq.~\eqref{eq:grav_action}, evaluated on the algebraic connection constraint, is the Einstein--Hilbert TT Hessian. Consequently $Q_T=M_{\rm Pl}^2$, $c_T^2=1$, there is no physical tensor mass term, and the two tensor polarizations have identical leading transfer functions. Appendix~A gives the model-specific TT response and auxiliary-field Hessian check underlying this conditional statement.

We use the exponential parametrization for the TT perturbation of the spatial metric
\begin{equation}
g_{ij}
=
a^{2}(t)\,[e^{h}]_{ij},
\qquad
\partial^{i}h_{ij}=0,
\qquad
h^{i}{}_{i}=0.
\label{eq:tt_perturbation}
\end{equation}
Because $\det(e^{h})=1$, the tensor-only synchronous path has $\sqrt{-g}=a^3$ and, for $\bar J_5^\mu=\rho_5\delta^\mu_0$ with TT-inert $\rho_5$, $\mathcal Q_{5,\rm MF}=-\rho_5^2$. The axial constraint then gives the response
\begin{equation}
\begin{aligned}
\delta_{\rm TT}A_\mu
&=-\frac{\kappa}{2(1-4\zeta)}\delta_{\rm TT}\bar J_{5\mu}=0,\\
\delta_{\rm TT}^{2}A_\mu
&=-\frac{\kappa}{2(1-4\zeta)}\delta_{\rm TT}^{2}\bar J_{5\mu}=0.
\end{aligned}
\label{eq:axial_tt_response}
\end{equation}
This vanishing concerns the axial irreducible component on the stated path, not the full connection: its Levi--Civita and tetrad-induced contortion responses are retained in the constrained variation. In the axial block, both the direct TT Hessian and the projected mixed block $S_{h_{\rm TT}A_\mu}$ vanish, so the axial Schur-complement term vanishes as well. Equivalently, the factorized contact action has zero first and second TT variations. A TT response of the connected term $\mathcal C_5$ would add a state-dependent matter contribution; this conclusion does not follow from one-point isotropy alone. Appendix~A supplies the full algebraic check.

Writing $T^{i}{}_{j}=p\delta^{i}{}_{j}+\pi^{i}{}_{j}$ with traceless TT anisotropic stress $\pi^{i}{}_{j}$, the quadratic tensor action on the homogeneous isotropic branch is the standard Einstein--Hilbert result,
\begin{equation}
S_{T}^{(2)}
=
\frac{M_{\rm Pl}^{2}}{8}
\sum_{\lambda=+,\times}
\int dt\,d^{3}x\,a^{3}
\left[
\dot h_{\lambda}^{2}
-
\frac{(\nabla h_{\lambda})^{2}}{a^{2}}
\right].
\label{eq:tensor_quadratic_action}
\end{equation}
Equivalently, the TT projection of the field equation is
\begin{equation}
\ddot h^{i}{}_{j}
+3H\dot h^{i}{}_{j}
-\frac{\nabla^{2}}{a^{2}}h^{i}{}_{j}
=
2\kappa\,\pi^{i}{}_{j}.
\label{eq:tensor_eom}
\end{equation}
We set $\pi^{i}{}_{j}=0$ on the branch studied here. The contact interaction may still modify the background $H(t)$ through the matter sector. A general normalization $Q_T(t)$ would add $\dot Q_T/Q_T$ to the tensor friction term, but here $Q_T=M_{\rm Pl}^{2}$ is constant: the usual $3H\dot h_{ij}$ remains with no additional propagation friction, kinetic renormalization, gradient correction, or physical tensor mass.

This establishes Proposition~1. It also clarifies why a ``frozen-torsion'' calculation can be misleading if the Cartan constraint is not enforced at the perturbed order. Holding an auxiliary torsion component fixed while varying only explicit metric factors does not, in general, remain on the connection equation and omits the mixed Hessian contribution in Eq.~\eqref{eq:schur_hessian}. Solving the algebraic connection equation first, or varying tetrad and connection consistently before imposing it, yields Eqs.~\eqref{eq:tensor_quadratic_action}--\eqref{eq:tensor_eom} instead of a mass-like tensor term.

\subsection{Dispersion Relation and Chromatic Propagation Test}
\label{sec:dispersion}

For a source-free subhorizon Fourier mode, Eq.~\eqref{eq:tensor_eom} gives
\begin{equation}
\omega^{2}=k_{\rm phys}^{2},\qquad k_{\rm phys}=k/a,
\label{eq:dispersion}
\end{equation}
so $v_p=v_g=c$, $\Delta v_g/c=0$, and there is no model-generated dispersive delay or phase accumulation: $\Delta t_{\rm disp}=\Delta\Phi_{\rm disp}=0$.

For comparison, a phenomenological mass-like template
\begin{equation}
\omega^{2}=k_{\rm phys}^{2}+\mu^{2}
\label{eq:massive_template}
\end{equation}
would give, in the relativistic regime,
\begin{equation}
\frac{\Delta v_{g}}{c}
\simeq
-\frac{\mu^{2}}{2(2\pi f)^{2}}
\propto f^{-2}.
\label{eq:massive_template_dv}
\end{equation}
Equation~\eqref{eq:massive_template_dv} is therefore a useful observational comparison template, but it is not generated by the algebraic quadratic-torsion operator in Eq.~\eqref{eq:grav_action}. A confirmed $f^{-2}$ propagation residual cannot be interpreted as evidence for the minimal $T^{2}$ Einstein--Cartan--Dirac sector without additional operators that change the physical tensor propagation operator.

The null result concerns the principal two-derivative characteristic. It does not imply that every torsion-dependent observable vanishes. In a geometric-optics expansion, the eikonal equation fixes the leading wave vector, whereas the next transport equation controls the evolution of amplitude and polarization. More general Riemann--Cartan backgrounds can therefore retain a null characteristic while acquiring torsion-dependent transport corrections, as found in weak-torsion analyses such as Ref.~\citep{Elizalde2023}. Evaluating the corresponding subprincipal transfer for the present Dirac-sourced algebraic branch would require perturbing the microscopic spin current consistently with the connection constraint. Such effects are not inserted phenomenologically into the leading dispersion relation here.

\subsection{Polarization Discriminants}
\label{sec:polarization}

The action in Eq.~\eqref{eq:grav_action} is parity even, although the isotropic ensemble of Eq.~\eqref{eq:axial_current} need not be parity symmetric. After the auxiliary torsion is eliminated, the leading TT Hessian contains no parity-odd tensor operator under the no-intrinsic-TT-response assumptions. Hence $\mathcal T_+(k,t)=\mathcal T_\times(k,t)$ and $A_+^{\rm obs}/A_\times^{\rm obs}=A_+^{\rm src}/A_\times^{\rm src}$ at leading order: the minimal model generates neither birefringence nor a differential polarization phase.

Chern--Simons gravity provides a useful qualitative contrast because a parity-violating interaction can distinguish the circular polarization states and produce birefringent propagation \citep{Alexander2009,Nishizawa2018}. A conventional massive-graviton template provides a second contrast: it can preserve polarization degeneracy while producing the mass-like $f^{-2}$ dispersion of Eq.~\eqref{eq:massive_template_dv}. The present algebraic EC--Dirac model predicts neither effect at leading order.

A robust propagation-induced birefringence signal would therefore be incompatible with the minimal homogeneous and isotropic realization considered here. Likewise, a confirmed chromatic propagation delay would not be described by the algebraic $T^{2}$ sector alone; examples of extensions that can modify the propagation operator include derivative torsion terms, propagating connection modes, additional curvature--torsion couplings, and other modified-gravity sectors.

\section{Observational Implications for Nanohertz Pulsar Timing Arrays}
\label{sec:numerical}

The action-level result of Sec.~\ref{sec:tensor} fixes the observational target before a likelihood is chosen. The algebraic $T^{2}$ sector predicts the same leading tensor operator as GR on the branch studied here, so the useful observables are departures from that baseline rather than a torsion-generated dispersive template. This section combines broadband propagation diagnostics, PTA spatial correlations, a covariance-aware estimator, and reported NANOGrav results while keeping the matter-sector role of $\zeta$ separate.

\subsection{Broadband and PTA Propagation Observables}
\label{sec:null_vector}
\label{sec:pta_null}
Relative to a GR wave on the same homogeneous background, define
\begin{equation}
\begin{aligned}
\boldsymbol{\mathcal O}_{\rm prop}&\equiv(\delta c_T,\delta_{\rm amp},\delta\Phi_{\rm disp},\Delta_{\rm pol}),\\
\boxed{\boldsymbol{\mathcal O}_{\rm prop}^{\rm EC}=(0,0,0,0)}.
\end{aligned}
\label{eq:observable_null}
\end{equation}
Here $\delta c_T$ measures the characteristic-speed shift, $\delta_{\rm amp}$ additional leading-order damping relative to GR on the same background, $\delta\Phi_{\rm disp}$ a chromatic phase, and $\Delta_{\rm pol}$ differential polarization transfer. Background-expansion changes and subprincipal spin transport are not included in these coordinates. Equation~\eqref{eq:observable_null} therefore classifies the entire leading propagation sector, not just one dispersion coefficient.

For a stochastic tensor background, the timing-residual cross spectrum of pulsars $a$ and $b$ can be written as
\begin{equation}
\left\langle \widetilde r_{a}(f)\widetilde r_{b}^{\ast}(f')\right\rangle
=
\frac{1}{2}\delta(f-f')\,\Gamma_{ab}(f)\,S_{r}(f),
\label{eq:pta_cross_spectrum}
\end{equation}
where $S_r(f)$ is the common residual spectrum and $\Gamma_{ab}$ the spatial kernel. Nonstandard phase velocity can make the overlap-reduction function (ORF) frequency dependent \citep{Liang2024PTA,Cordes2025PTA,Inomata2024ORF}; massive-graviton analyses have already used PTA angular-correlation information to constrain such dispersive propagation \citep{WangZhao2024}. By contrast, the luminal and nondispersive EC--Dirac tensor operator gives $\Gamma_{ab}^{\rm EC}(f)=\Gamma_{ab}^{\rm GR}(f)$ for the same source model, geometry, and pulsar-term treatment.

\begin{figure*}
    \centering
    \includegraphics[width=0.90\linewidth]{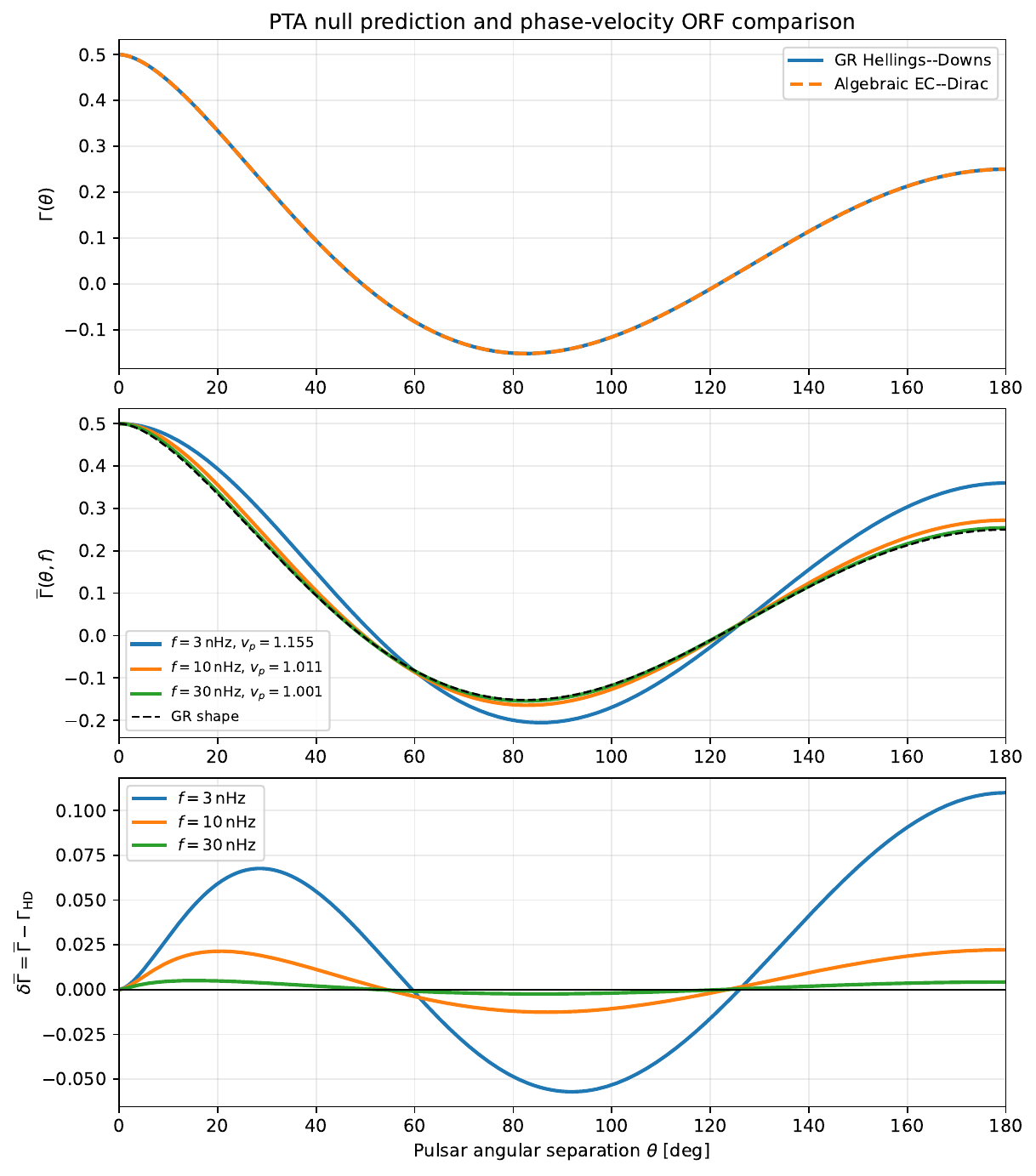}
    \caption{Analytic PTA comparison generated from Eqs.~\eqref{eq:hd_curve}--\eqref{eq:phase_velocity_response}. Top: the algebraic EC--Dirac prediction coincides with the GR Hellings--Downs curve. Middle: a shape-normalized nonstandard-phase-velocity ORF for $f=3$, $10$, and $30\,$nHz with the illustrative scale $f_m=1.5\,$nHz. Bottom: the corresponding shape residual $\delta\overline\Gamma$. No timing data enter this figure; the non-GR curves are comparison templates based on the ORF framework of Refs.~\citep{Liang2024PTA,Cordes2025PTA}.}
    \label{fig:hd_null}
\end{figure*}

For distinct pulsars in the ideal isotropic, unpolarized, short-wavelength tensor limit, the GR kernel is the Hellings--Downs curve \citep{HellingsDowns1983},
\begin{equation}
\Gamma_{\rm HD}(\theta)
=
\frac{1}{2}+\frac{3}{2}x\ln x-\frac{1}{4}x,
\qquad x\equiv\frac{1-\cos\theta}{2}.
\label{eq:hd_curve}
\end{equation}
A purely frequency-dependent rescaling $g(f)\Gamma_{\rm HD}(\theta)$ is degenerate with $S_r(f)$ and is not, by itself, an identifiable deformation of the spatial kernel. We therefore define a shape residual
\begin{samepage}
\begin{equation}
\begin{aligned}
\overline{\Gamma}(\theta,f)&\equiv\frac{\Gamma(\theta,f)}{2\Gamma(0,f)},\\
\delta\overline{\Gamma}(\theta,f)&\equiv\overline{\Gamma}(\theta,f)-\Gamma_{\rm HD}(\theta).
\end{aligned}
\label{eq:shape_normalized_orf}
\end{equation}
\end{samepage}
so that $\overline{\Gamma}(0,f)=1/2$ at every frequency. For illustration only, Fig.~\ref{fig:hd_null} uses the nonstandard phase-velocity response
\begin{equation}
\begin{aligned}
F_a^A(\widehat\Omega;v_p)&=
\frac{v_p\,\widehat p_a^i\widehat p_a^j e_{ij}^A(\widehat\Omega)}
{2[v_p+\widehat\Omega\!\cdot\!\widehat p_a]},\\
v_p(f)&=\left[1-\left(\frac{f_m}{f}\right)^2\right]^{-1/2}.
\end{aligned}
\label{eq:phase_velocity_response}
\end{equation}
with $f_m=1.5\,$nHz. This comparison template is not generated by the EC--Dirac action; it visualizes the kind of angular-shape change that a frequency-resolved PTA likelihood can identify. Finite-distance, autocorrelation, and pulsar-term effects require the full response treatment \citep{Cordes2025PTA}.

\subsection{Covariance-Weighted Statistic for a Future PTA Analysis}
\label{sec:pta_stat}
\label{sec:forecast}
Let $\alpha=(ab,i)$ label a pulsar pair and frequency bin, $d_\alpha\equiv\widehat\Gamma_{ab}(f_i)-\Gamma_{ab}^{\rm GR}(f_i)$, and $\mathbf C$ be the covariance after timing and common-spectrum nuisance parameters are treated consistently. For an identifiable shape template $(t_\lambda)_\alpha=K_{ab}(f_i;\lambda)$, a local amplitude is
\begin{equation}
\widehat A_\lambda=
\frac{\mathbf t_\lambda^{\rm T}\mathbf C^{-1}\mathbf d}
{\mathbf t_\lambda^{\rm T}\mathbf C^{-1}\mathbf t_\lambda},
\qquad
\sigma^{-2}(A_\lambda)=\mathbf t_\lambda^{\rm T}\mathbf C^{-1}\mathbf t_\lambda.
\label{eq:pta_gls_estimator}
\end{equation}
The minimal theory is the nested zero-deformation case $\mathcal H_{\rm EC}:A_\lambda=0$ against an extended propagation model with $A_\lambda$ free. This form is compatible with present work on frequency-dependent ORFs, alternative polarizations, and more general PTA response functions \citep{Liang2024PTA,Cordes2025PTA,Hu2025Pol,Inomata2024ORF}. A theory-agnostic score $\mathbf d^{\rm T}\mathbf C^{-1}\mathbf d$ must be calibrated under the full timing likelihood rather than assigned a universal $\chi^2$ distribution.

\begin{figure*}[t]
\centering
\includegraphics[width=0.90\linewidth]{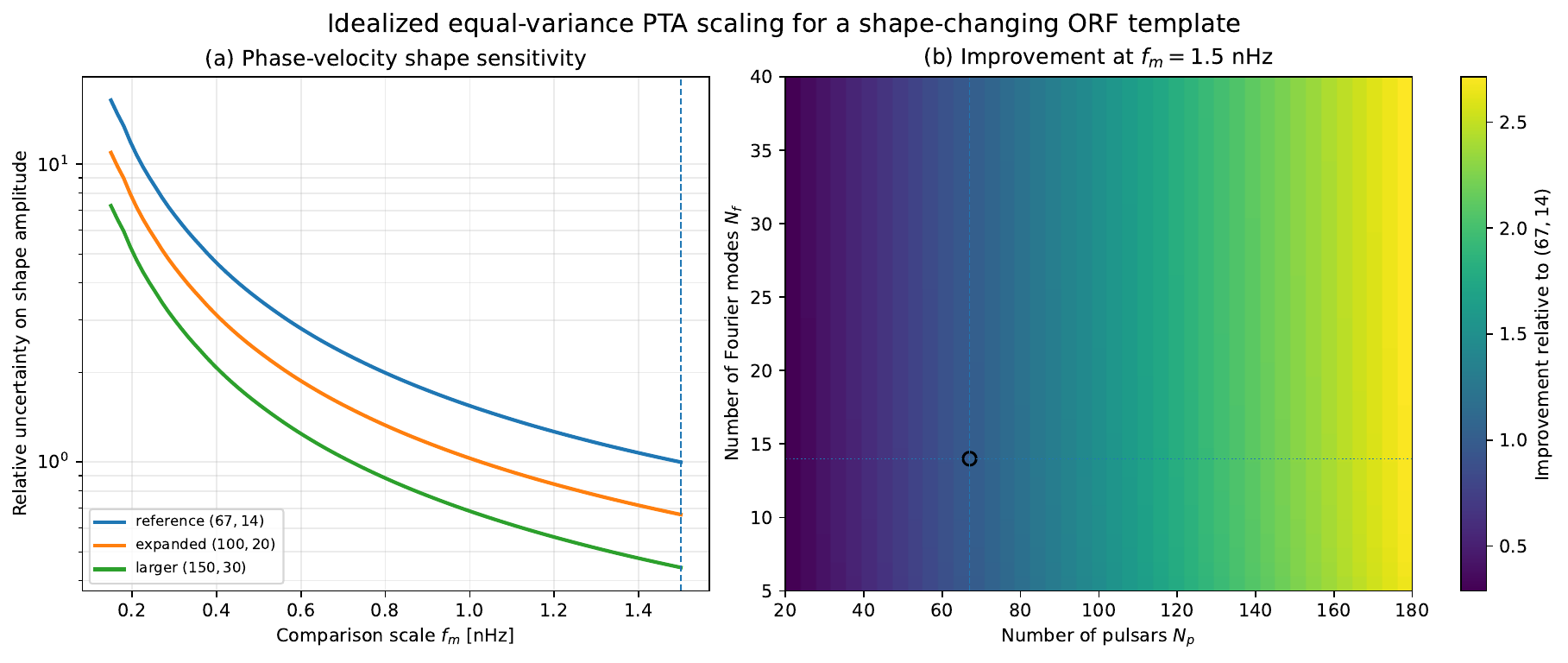}
\caption{Idealized equal-variance scaling for the shape-changing ORF template in Fig.~\ref{fig:hd_null}. Left: relative uncertainty as the comparison scale $f_m$ varies. Right: information gain at $f_m=1.5\,$nHz as pulsar and frequency-bin counts increase. The open circle marks the $(67,14)$ count-scale reference. The curves show pair-count and frequency leverage only; no timing residuals or posterior samples enter the calculation.}
\label{fig:forecast}
\end{figure*}

The statistic in Eq.~\eqref{eq:pta_gls_estimator} is the appropriate object for a dataset-level implementation because the timing model, common spectrum, pulsar-specific noise, pulsar distances, and cross-pulsar covariance can all enter through $\mathbf C$ and the response template. For illustration only, assume independent pair--frequency estimates with common variance $\sigma_\Gamma^2$, and set $K_{ab}=\delta\overline\Gamma(\theta_{ab},f_i;f_m)$. For isotropically distributed distinct pairs, define
\begin{equation}
\begin{aligned}
\mathcal J(f_i;f_m)&\equiv\int_0^1[\delta\overline\Gamma(x,f_i;f_m)]^2dx,\\
\left\langle\sigma^{-2}(A)\right\rangle_{\rm iso}
&=\frac{N_p(N_p-1)}{2\sigma_\Gamma^2}\sum_{i=1}^{N_f}\mathcal J(f_i;f_m).
\end{aligned}
\label{eq:forecast_sigma}
\end{equation}
Figure~\ref{fig:forecast} illustrates this pair-count and frequency-bin scaling, using $(N_p,N_f)=(67,14)$ only as a count-scale reference to the published NANOGrav 15-year analysis \citep{NANOGrav15yr}. The approximation omits shared-pulsar covariance, unequal noise, pulsar distances, cadence, chromatic noise, and timing-model parameters; it is not a collaboration forecast or a projected constraint. Flexible spectral refitting and chromatic-noise models can affect a full inference \citep{Esmyol2026,Larsen2026Chromatic,NANOGravChromatic2026}.

\subsection{Consistency with NANOGrav 15-year Results}
\label{sec:ng15_data}
The zero-deformation prediction can be checked against published NANOGrav angular-correlation posteriors. We use the published posterior summaries as an empirical baseline; we do not reconstruct the collaboration timing likelihood or infer a new posterior for $\zeta$. The NANOGrav harmonic analysis expands the pulsar-pair correlations into Legendre multipoles and, for a quadrupole-only model, reports \citep{NANOGravHarmonic2025}
\begin{equation}
R_2\equiv\frac{c_2}{c_2^{\rm HD}}=1.088^{+0.32}_{-0.45}
\qquad(68\%\ \text{credible interval}),
\label{eq:ng15_r2}
\end{equation}
with a Bayes factor of about 200 relative to an uncorrelated common-red-noise model. The EC--Dirac prediction is $R_2=1$, or $\Delta_2\equiv R_2-1=0$, which lies inside the published 68\% interval.

\begin{figure*}
    \centering
    \includegraphics[width=0.92\linewidth]{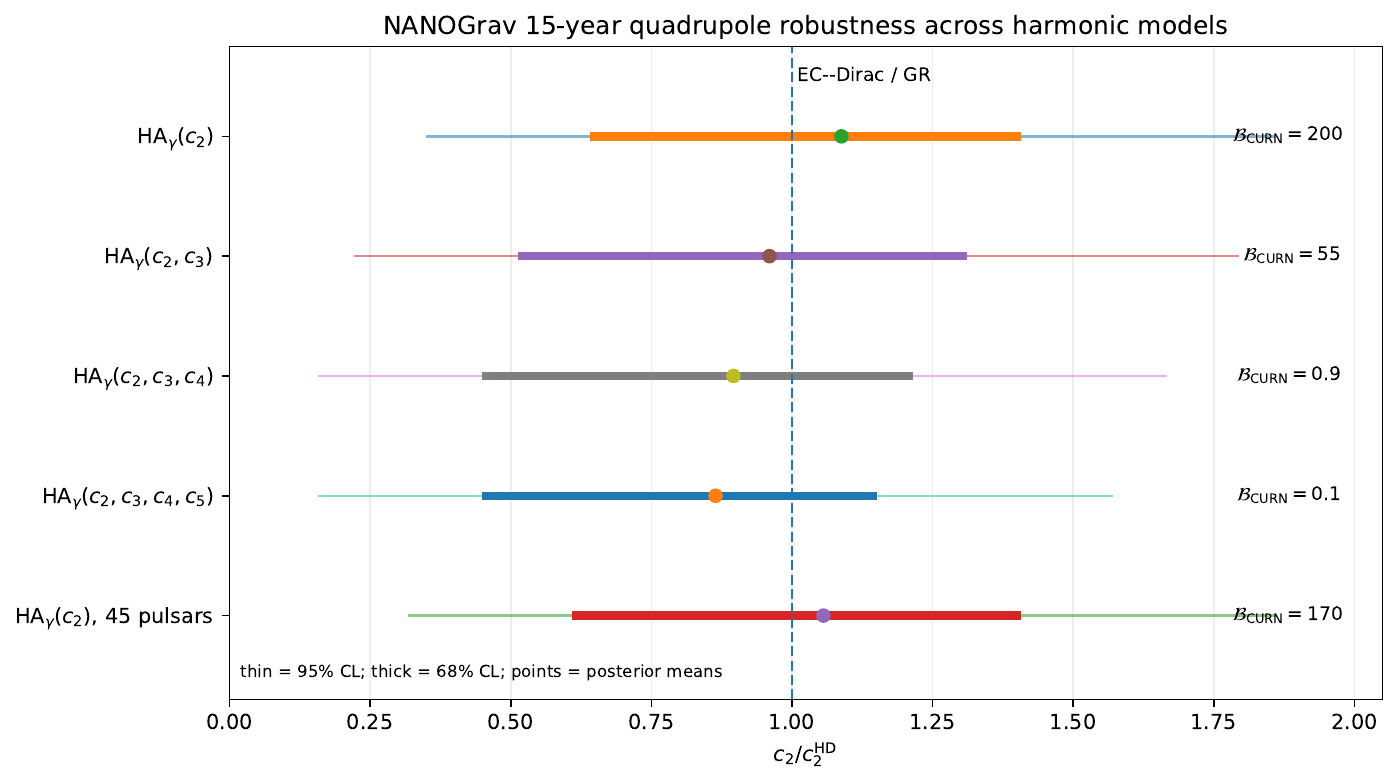}
    \caption{Published NANOGrav 15-year harmonic-model summaries compared with the shared GR/EC--Dirac quadrupole value \citep{NANOGravHarmonic2025}. Points show the reported posterior means of $c_2/c_2^{\rm HD}$; thick and thin intervals are the published 68\% and 95\% credible intervals, respectively. The vertical dashed line marks the value $1$. Right-hand annotations give the published Bayes factors of the full harmonic model relative to an uncorrelated common-red-noise model. All listed 68\% intervals contain $1$. These summaries provide a consistency check shared with GR; they do not identify torsion or constrain $\zeta$. No collaboration likelihood is reconstructed here.}
    \label{fig:ng15_harmonic}
\end{figure*}

A stronger robustness check is available from the same published analysis. Its Table~3 gives the marginalized $c_2$ posterior after successively freeing $c_3$, $c_4$, and $c_5$, and after restricting the data to the 45 pulsars inherited from the 12.5-year sample. Figure~\ref{fig:ng15_harmonic} rescales those reported intervals by $c_2^{\rm HD}=0.3125$. In every listed harmonic truncation, and in the 45-pulsar subset, the value $R_2=1$ remains inside the published 68\% interval. The posterior mean shifts from $R_2=1.088$ for $\mathrm{HA}_\gamma(c_2)$ to $0.864$ when multipoles through $\ell=5$ are freed, but this drift is smaller than the corresponding uncertainty. The decreasing Bayes factors for models with additional multipoles should not be read as a probability for the EC--Dirac point: they compare the full higher-dimensional harmonic models with the uncorrelated common process and primarily reflect the absence of evidence for the extra multipoles. The 45-pulsar subset retains a Bayes factor of about 170 and a quadrupole mean $R_2=1.056$, showing that the GR-like quadrupole is not driven by the additional short-baseline pulsars.

When monopole and dipole components are also allowed, the quadrupole evidence is reduced by a low-frequency monopolar feature near $4\,$nHz; this is an important nuisance structure for future frequency-resolved tests and prevents overinterpreting the one-dimensional comparison. Posterior-predictive checks of the same 15-year data find the interpulsar correlations and spectral coefficients statistically consistent with a Hellings--Downs/power-law description \citep{NANOGravPPC2025}.

\paragraph{2026 collaboration preprint: full-timing likelihood robustness.}
The 2026 NANOGrav collaboration preprint (arXiv:2606.28554) reports updated full-timing results for the 67-pulsar GWB sample using customized chromatic-noise models \citep{Larsen2026Chromatic,NANOGravChromatic2026}. With a 14-mode power-law common process it reports ${\rm BF}_{\rm HD/CURN}=1571\pm14$ and an interpulsar-correlation significance of $3.32\sigma$. We quote these collaboration-level likelihood results directly rather than present a logarithmic conversion as an independent evidence calculation. The same reanalysis raises the quoted correlation significance from $3.16\sigma$ to $3.32\sigma$ and shifts the fixed-$\gamma_{\rm GWB}=13/3$ strain amplitude from $2.4^{+0.7}_{-0.6}\times10^{-15}$ in the original analysis to $2.1^{+0.6}_{-0.5}\times10^{-15}$ \citep{NANOGrav15yr,NANOGravChromatic2026}. In the varied-index analysis the reported median moves from $\gamma_{\rm GWB}=3.2$ to $3.5$. These spectral changes do not alter the EC--Dirac propagation statement because the leading prediction concerns the normalized spatial kernel rather than the source-spectrum amplitude or slope.

The customized-noise census is particularly relevant for a future chromatic propagation analysis: non-dispersive chromatic delays are identified in $21/67=31.3\%$ of pulsars, while the inferred achromatic-noise component changes materially in $19/67=28.4\%$ \citep{Larsen2026Chromatic}. The 2026 collaboration preprint reports that, after incorporating those pulsar-specific chromatic models, the evidence for the Hellings--Downs spatial pattern increases rather than decreases. The reported quantities are summarized in Table~\ref{tab:ng15_likelihood_summary}; the values in parentheses are reported credible ranges, not uncertainties inferred in this work.

\begin{table*}[t]
\centering
\footnotesize
\caption{Published NANOGrav 15-year and 2026 collaboration-preprint likelihood summaries. The values are quoted from the cited sources; this work performs no reanalysis and obtains no torsion-parameter constraint.}
\label{tab:ng15_likelihood_summary}
\begin{tabular}{lcc}
\hline
Quantity & 2023 analysis & 2026 customized-CNM analysis \\
\hline
${\rm BF}_{\rm HD/CURN}$ & $200$--$1000$ & $1571\pm14$ \\
Interpulsar correlation & $3.16\sigma$ (like-for-like baseline) & $3.32\sigma$ \\
$A_{\rm GWB}$, $\gamma=13/3$ & $2.4^{+0.7}_{-0.6}\times10^{-15}$ & $2.1^{+0.6}_{-0.5}\times10^{-15}$ \\
Varied $\gamma_{\rm GWB}$ & $3.2^{+0.6}_{-0.6}$ & $3.5^{+0.7}_{-0.6}$ \\
\hline
\end{tabular}\par
\vspace{2pt}
\emph{Source:} Refs.~\citep{NANOGrav15yr,Larsen2026Chromatic,NANOGravChromatic2026}; 2023 values are from the published analysis and 2026 full-timing values from the collaboration preprint.
\end{table*}
The current EC--Dirac branch predicts exactly the HD spatial kernel at leading order. The latest nuisance-model update therefore strengthens the empirical baseline against which a future frequency-dependent shape deformation must be tested, but it does not preferentially select the EC--Dirac model over GR and does not provide a direct bound on $\zeta$.

The latest public timing release is NANOGrav 15-year v2.1.0, which adds narrowband post-fit residual products \citep{NANOGravTimingV21}. For free-spectrum work, NANOGrav released corrected KDE representations in August 2026 after identifying a parallel-tempering sampler issue in older free-spectrum chains \citep{NANOGravKDE2026}. A separate 2026 likelihood study also finds that improving the standard Fourier approximation changes the absolute marginal likelihoods for power-law processes but affects the HD/CURN comparison very similarly, leaving that model comparison essentially unchanged \citep{Olum2026Fourier}. We therefore treat the corrected public products and the 2026 preprint likelihood as the appropriate current public baseline. These real-data checks motivate, rather than replace, a dedicated TOA-level fit of $\delta\overline\Gamma(\theta,f)$ with the complete public timing likelihood.

\subsection{Polarization and Multi-Band Observational Probes}
\label{sec:polarization_null}
\label{sec:constraints}
The minimal branch also predicts $\Delta_A^{\rm pol}\equiv\ln|\mathcal T_+/\mathcal T_\times|=0$ and $\Delta_\phi^{\rm pol}\equiv\arg\mathcal T_+-\arg\mathcal T_\times=0$. PTA analyses with alternative polarization templates already provide a direct observational framework for this channel \citep{Chu2021,Kumar2024,Hu2025Pol}, while newer multi-point strategies offer complementary tests of non-tensorial polarizations \citep{JimenezCruz2026}. A measured polarized signal would still have to be separated from source-population polarization before it could be interpreted as propagation physics.

The same propagation law can be tested across widely separated frequencies without requiring identical source populations. GW170817/GRB~170817A constrains the tensor characteristic to be extremely close to luminal at ground-based frequencies \citep{GW170817,Baker2017scalar}; the present model satisfies this identically at leading order and therefore receives no numerical bound on $\zeta$ from that event. At nanohertz frequencies, the absence of a model-generated mass term means that lowering $f$ does not uncover a hidden $f^{-2}$ coefficient. Instead, PTAs test the angular and polarization response directly. This multi-band logic is consistent with the broader role of PTAs as probes of fundamental gravity reviewed in Ref.~\citep{Taylor2025PTAReview}. Table~\ref{tab:null_channels} summarizes the leading propagation channels, while Table~\ref{tab:empirical_status} connects them to current observational baselines.

\begin{table*}[t]
\centering
\footnotesize
\caption{Theory-derived leading propagation channels of the minimal algebraic EC--Dirac branch. The entries follow from Sec.~\ref{sec:tensor}; no observational data are used in this table.}
\label{tab:null_channels}
\begin{tabular}{>{\raggedright\arraybackslash}p{0.26\linewidth}>{\centering\arraybackslash}p{0.16\linewidth}>{\raggedright\arraybackslash}p{0.48\linewidth}}
\hline
Channel & Prediction & Interpretation of a robust nonzero contribution \\
\hline
Tensor speed $\delta c_T$ & $0$ & Modified principal tensor operator \\
Chromatic phase/delay & $0$ & Dispersive or derivative propagation sector \\
PTA shape residual $\delta\overline\Gamma$ & $0$ & Frequency-dependent spatial kernel beyond the GR baseline \\
Polarization transfer $\Delta_{A,\phi}^{\rm pol}$ & $0$ & Parity-odd, anisotropic, or extra connection structure \\
Extra amplitude damping & $0$ & Running tensor normalization or another nonminimal operator \\
\hline
\end{tabular}
\end{table*}

\begin{table*}[t]
\centering
\footnotesize
\caption{Current empirical status. Numerical statements are taken from the cited observational papers; the final column gives their interpretation for the present model.}
\label{tab:empirical_status}
\begin{tabular}{>{\raggedright\arraybackslash}p{0.18\linewidth}>{\raggedright\arraybackslash}p{0.35\linewidth}>{\raggedright\arraybackslash}p{0.37\linewidth}}
\hline
Probe & Empirical status & Implication for this model \\
\hline
GW170817 & Tensor propagation is constrained to be extremely close to luminal at ground-based frequencies \citep{GW170817,Baker2017scalar}. & Satisfied identically at leading order; consistency check, not a direct bound on $\zeta$. \\
PTA spatial correlations & NANOGrav, EPTA+InPTA, and CPTA report a nanohertz common process and/or correlations compatible with an isotropic tensor background \citep{NANOGrav15yr,EPTA2023,CPTA2023}. NANOGrav gives Eq.~\eqref{eq:ng15_r2} \citep{NANOGravHarmonic2025}. & The predicted ratio $1$ lies inside the published interval; current data are consistent with the zero-deformation propagation class. \\
Chromatic/polarization PTA channels & Frequency-dependent ORFs, alternative-polarization statistics, and refined noise models are available \citep{Liang2024PTA,Cordes2025PTA,Hu2025Pol,Larsen2026Chromatic}. & These analyses provide the machinery for a frequency-resolved test of $A_\lambda=0$ and $\Delta_{A,\phi}^{\rm pol}=0$. \\
CMB/BAO/SNe & A specified torsion cosmology has been constrained with DESI, supernovae, and CMB, yielding a model-specific torsion parameter consistent with zero \citep{Liu2025Torsion}. & The fitted cosmological parameter is not automatically the $\zeta$ of Eq.~\eqref{eq:grav_action}; a mapping of the action and matter closure is required. \\
\hline
\end{tabular}
\end{table*}

\subsection{Matter-Sector Regularity and Perturbative Control}
\label{sec:benchmark}
Although $\zeta$ drops out of leading GW propagation, it remains in the axial--axial contact interaction. Invertibility of the full algebraic connection quadratic form requires $\zeta\notin\left\{-\frac12,\frac14,2\right\}$. A dimensional perturbative-control criterion for the reduced four-fermion operator is $|C_{AA}|E^2\lesssim1$, where $|C_{AA}|=3\kappa/[16|1-4\zeta|]$. At the illustrative energy scale $E=10^{12}\,$GeV this criterion gives $|1-4\zeta|\gtrsim3.2\times10^{-14}$. It estimates the strong-coupling region around $\zeta=1/4$; it is not an observational constraint.

Under the factorized closure, one may define $|C_{AA}|\rho_5^2$ as a dimensional interaction scale. It is not, by itself, the physical energy density of a specified fermionic state. We therefore do not compare it with the present critical density or infer a cosmological bound from it. A physical matter-sector constraint requires the state-dependent stress tensor, the composite-current correlations, and their cosmological evolution.

A recent DESI/SNe/CMB torsion-cosmology fit finds a model-specific parameter $\alpha=-0.00066\pm0.00098$ \citep{Liu2025Torsion}. We quote this only as an example of a constraint within a specified cosmological model; no model-independent map from that $\alpha$ to the present $\zeta$ exists. The regularity conditions and perturbative-control estimate above do not supply an observational posterior for $\zeta$.

For reproducibility, Figs.~\ref{fig:hd_null} and \ref{fig:forecast} are generated from the analytic equations and inputs stated here. Figure~\ref{fig:ng15_harmonic} uses the published NANOGrav summary in Eq.~\eqref{eq:ng15_r2}; the tables distinguish theory summaries from observational results and cite their sources.

\section{Discussion}
\label{sec:discussion}

\subsection{Physical Interpretation and Discriminating Power}
The physical result is a separation between two roles of torsion. The quadratic coupling survives in the reduced Dirac matter interaction, but it does not enter the leading TT kinetic, gradient, mass, or polarization-splitting operators once the auxiliary Lorentz connection is eliminated consistently. The relevant chain is first-order EC--Dirac theory $\rightarrow$ algebraic connection reduction $\rightarrow$ constrained TT Hessian $\rightarrow$ observable propagation coordinates. The TT-response and projected Schur-complement argument in Sec.~\ref{sec:tensor_action}, supported by the detailed convention check in Appendix~A, explains why holding an auxiliary torsion component fixed during a metric variation can generate a spurious mass-like term.

The leading propagation prediction is identical to the GR prediction under the assumptions of Proposition~1. Agreement of PTA data with the Hellings--Downs response is therefore a consistency check and cannot distinguish this branch from GR. A statistically robust non-GR propagation signal, after accounting for pulsar and chromatic-noise systematics, could disfavor the minimal branch considered here. Such a signal would not by itself identify torsion; its interpretation would require comparison with other propagation mechanisms.

\subsection{Relation to Recent Torsion and PTA Studies}
Table~\ref{tab:theory_comparison} places the present conditional calculation alongside general results on nondynamical torsion, model-specific Einstein--Cartan perturbation studies, and recent PTA analyses.

\begin{table*}[t]
\centering
\scriptsize
\caption{Relation to recent work. The table emphasizes differences in dynamical sector, observable, or inference target rather than ranking the studies.}
\label{tab:theory_comparison}
\begin{tabular}{>{\raggedright\arraybackslash}p{0.23\linewidth}>{\raggedright\arraybackslash}p{0.31\linewidth}>{\raggedright\arraybackslash}p{0.38\linewidth}}
\hline
Study & Main target & Relation to the present work \\
\hline
Karananas et al.~\citep{PhysRevD.104.064036} & Gravitational spectra with nondynamical torsion & Establishes a massless-graviton result for a broad model class; here we specialize to a Dirac-sourced $T^2$ invariant and an explicit conditional TT response. \\
Elizalde et al.~\citep{Elizalde2023} & Weak-torsion GW amplitude and polarization transport & Complementary subprincipal effects; here we derive the leading constrained TT operator. \\
Ranjbar et al.~\citep{Ranjbar2024} & Scalar/tensor perturbations in EC cosmology & Spin-fluid evolution, rather than the Dirac auxiliary-torsion and PTA analysis here. \\
Wang and Zhao~\citep{WangZhao2024} & PTA limits on massive-graviton dispersion & Fits nonzero dispersion; the minimal EC--Dirac branch predicts zero deformation. \\
Liang et al.~\citep{Liang2024PTA}; Cordes et al.~\citep{Cordes2025PTA} & Frequency-dependent ORFs for nonstandard phase velocity & Supplies comparison templates and tools for future likelihoods. \\
Hu et al.~\citep{Hu2025Pol} & Alternative polarizations in NANOGrav 15-year data & Tests the channel for which this model predicts no propagation-induced splitting. \\
NANOGrav analyses \citep{NANOGravHarmonic2025,NANOGravPPC2025,Larsen2026Chromatic,NANOGravChromatic2026} & Harmonic correlations, posterior checks, chromatic-noise systematics & Establishes the GR-like baseline and noise treatment for a dedicated test. \\
Liu et al.~\citep{Liu2025Torsion} & DESI/SNe/CMB fit of a specified torsion cosmology & Model-specific; no automatic map to the present $\zeta$. \\
Esmyol et al.~\citep{Esmyol2026}; Zhou and Li~\citep{ZhouLi2026} & Spectral refits and anisotropic PTA ORFs & Methods to extend the present isotropic test. \\
\hline
\end{tabular}
\end{table*}

A related action-level ordering appears in dynamical Barbero--Immirzi gravity \citep{Gao2026BI}, but the physical content is different: that analysis contains a dynamical pseudoscalar/nonminimal curvature sector and targets tensor normalization and standard-siren amplitudes, whereas the present coupling $\zeta$ survives only in the Dirac contact interaction and the observational matching is to PTA angular, chromatic, and polarization tests. More general modified-gravity strategies, including higher-order correlation functions, further illustrate how independent observables can separate tensor from non-tensor sectors \citep{JimenezCruz2026}.

\subsection{Limitations and Outlook}
Four limitations remain explicit. First, $\rho_5$ is a coarse-grained one-point quantity. A predictive spinor cosmology must specify the microscopic state, composite-current correlations, equation of state, and perturbations. Second, Proposition~1 is conditional on factorizing the axial-current contact operator, neglecting the connected term $\mathcal C_5$, imposing zero first- and second-order TT response of the factorized scalar, and setting the intrinsic TT anisotropic stress to zero. Statistical isotropy of $\langle J_5^\mu\rangle$ alone implies none of these composite-response conditions. A generic quantum or statistical state can have a nonzero connected two-point function whose TT response would contribute to the tensor Hessian; that response requires a specified microscopic state. Third, the propagation statement is classical, tree level, and leading two derivative. Subprincipal Riemann--Cartan transport, radiative corrections, or derivative/propagating torsion operators lie outside Eq.~\eqref{eq:grav_action} \citep{Shapiro2002,Elizalde2023}. Fourth, the NANOGrav comparison uses published harmonic posteriors and likelihood results reported in a 2026 collaboration preprint rather than a rerun of the timing likelihood, so it does not yet constrain the frequency-dependent comparison scale $f_m$.

The remaining next step is more specific: fit the shape residual $\delta\overline\Gamma(\theta,f)$ directly to the released TOAs with the collaboration-level chromatic-noise, ephemeris, pulsar-distance, red-noise, and cross-pulsar covariance structure. The latest chromatic-noise work shows why this nuisance treatment is essential \citep{Larsen2026Chromatic,NANOGravChromatic2026}. A second extension is to relax isotropy using analytic anisotropic ORFs \citep{ZhouLi2026}; a third is to build a self-consistent Einstein--Dirac cosmology so that matter-sector data can constrain $\zeta$ rather than a phenomenological normalization.

\section{Conclusion}
\label{sec:conclusion}
This work derives the leading tensor response of a quadratic Einstein--Cartan--Dirac model after eliminating its nondynamical Lorentz connection. The torsion coupling remains in the axial--axial matter interaction, while the leading tensor propagation is GR-like under the stated mean-field and TT-response closure. This is a conditional, model-specific result: the closure is not derived from one-point isotropy or from a microscopic fermionic state, and the calculation does not establish a general absence of torsion effects.

The comparison with published PTA summaries is a consistency check shared with GR. It does not identify algebraic torsion or constrain its coupling, and no new timing-data likelihood is evaluated. A robust departure from the GR propagation pattern could disfavor the minimal branch, but would not by itself identify torsion as its cause.

The next step is a state-resolved Einstein--Dirac calculation that determines the axial-current response and matter stress tensor self-consistently. With that input, a future PTA analysis can fit the timing data and relevant noise systematics directly, testing whether any observed propagation departure is compatible with this model.

\section*{Data Availability}
Inputs are limited to equations stated here and cited public NANOGrav summaries and releases, including the timing release v2.1.0 \citep{NANOGravTimingV21} and corrected August 2026 KDE release \citep{NANOGravKDE2026}. These data are not redistributed. No private data, TOA-level likelihood, or new MCMC analysis is used.

\section*{Declaration of Competing Interest}
The authors declare that they have no known competing financial
interests or personal relationships that could have appeared to
influence the work reported in this paper.

\section*{Funding}
This work was supported by the National Natural Science Foundation of
China (Nos.~12288102, 12573052, and 12475046) and the Major Science
and Technology Special Project of Xinjiang Uygur Autonomous Region
(No.~2022A03013-3).

\section*{Author Contributions}
Zhi-Fu Gao: Conceptualization, Methodology, Formal analysis, Writing -- original draft, Writing -- review and editing. Biao-Peng Li: Methodology, Formal analysis, Investigation, Writing -- review and editing. Na Wang: Supervision, Validation, Writing -- review and editing. Hui Wang: Conceptualization, Supervision, Funding acquisition, Writing -- review and editing. Guo-Qiang Jin: Formal analysis, Investigation. Zhoujian Cao: Formal analysis, Writing -- review and editing. Luiz C. Garcia de Andrade: Conceptualization, Formal analysis, Validation, Writing -- review and editing.

\appendix
\setcounter{figure}{0}
\setcounter{table}{0}
\renewcommand{\thefigure}{\Alph{section}.\arabic{figure}}
\renewcommand{\thetable}{\Alph{section}.\arabic{table}}

\section{Axial Reduction and Convention Check}
\label{app:variation}

This appendix records the algebraic reduction used in Sec.~\ref{subsec:action} and fixes the normalization of the quadratic torsion coupling. The orientation, gamma-matrix, and spinor-derivative conventions are those of Eqs.~\eqref{eq:axial_defs}--\eqref{eq:spinor_conventions}. Following the standard Lorentz-irreducible decomposition \citep{Hehl1976,Shapiro2002}, define the trace vector $V_\mu\equiv T^{\lambda}{}_{\mu\lambda}$, the axial vector $A^\mu$ of Eq.~\eqref{eq:axial_defs}, and a traceless tensor $Q_{\lambda\mu\nu}$ satisfying $Q^{\lambda}{}_{\mu\lambda}=0$ and $\epsilon^{\mu\nu\rho\sigma}Q_{\nu\rho\sigma}=0$. Up to the total divergence $-2\nabla_\mu V^\mu$, the two invariants entering Eq.~\eqref{eq:grav_action} decompose as
\begin{align}
R(\omega)
&=
R(g)-\frac{4}{3}V_\mu V^\mu
+\frac{3}{2}A_\mu A^\mu
+\frac{1}{2}Q_{\lambda\mu\nu}Q^{\lambda\mu\nu},
\label{eq:app_R_decomp}\\
T_{\lambda\mu\nu}T^{\lambda\mu\nu}
&=
\frac{2}{3}V_\mu V^\mu
-6A_\mu A^\mu
+Q_{\lambda\mu\nu}Q^{\lambda\mu\nu}.
\label{eq:app_T2_decomp}
\end{align}
Hence the algebraic gravitational quadratic form is proportional to
\begin{equation}
\frac{2}{3}(\zeta-2)V^2
+\frac{3}{2}(1-4\zeta)A^2
+\frac{1}{2}(1+2\zeta)Q^2,
\label{eq:app_full_quadratic_form}
\end{equation}
which makes the exceptional values $\zeta=2$, $1/4$, and $-1/2$ manifest. At $\zeta=2$ and $\zeta=-1/2$ the unsourced trace or tensor quadratic coefficient vanishes, respectively, so the corresponding auxiliary connection equation loses algebraic invertibility. These are flat directions of the algebraic connection sector, not new propagating modes, because no derivative kinetic term for these components is present in Eq.~\eqref{eq:grav_action}. Minimal Dirac matter sources only the axial piece, so the part of the first-order Lagrangian relevant for the sourced connection equation is
\begin{equation}
\mathcal L_A
=
\frac{3(1-4\zeta)}{4\kappa}A_\mu A^\mu
+
\frac{3}{4}A_\mu J_5^\mu .
\label{eq:app_LA}
\end{equation}
Varying Eq.~\eqref{eq:app_LA} gives $[3(1-4\zeta)/(2\kappa)]A_\mu+(3/4)J_{5\mu}=0$. On the regular axial branch,
\begin{equation}
A_\mu=-\frac{\kappa}{2(1-4\zeta)}J_{5\mu},
\label{eq:app_A_solution}
\end{equation}
and substitution gives
\begin{equation}
\mathcal L_A^{\rm red}
=-\frac{3\kappa}{16(1-4\zeta)}J_{5\mu}J_5^\mu.
\label{eq:app_LA_reduced}
\end{equation}
At $\zeta\to0$, Eq.~\eqref{eq:app_LA_reduced} reduces to the standard EC--Dirac contact term in our conventions. At $\zeta=1/4$ the sourced algebraic solution ceases to be regular; the other exceptional values arise only from the unsourced trace and tensor sectors. None represents a new propagating mode.

\subsection{Hessian-level consistency of the auxiliary-field reduction}
Let $\Phi=(e,\psi)$ denote the fields retained after the Lorentz connection is eliminated, and let $\omega_\star(\Phi)$ solve $S_\omega(\Phi,\omega_\star)=0$ on the regular branch. Differentiating the connection equation with respect to $\Phi$ gives
\begin{equation}
S_{\omega\Phi}+S_{\omega\omega}\,\omega_{\star,\Phi}=0,
\qquad
\omega_{\star,\Phi}=-S_{\omega\omega}^{-1}S_{\omega\Phi}.
\label{eq:app_aux_response}
\end{equation}
For $S_{\rm eff}(\Phi)=S[\Phi,\omega_\star(\Phi)]$, a second derivative therefore yields
\begin{equation}
S_{{\rm eff},\Phi\Phi}
=
S_{\Phi\Phi}-S_{\Phi\omega}S_{\omega\omega}^{-1}S_{\omega\Phi},
\label{eq:app_schur_proof}
\end{equation}
which is Eq.~\eqref{eq:schur_hessian}. Thus integrating out the algebraic connection before expanding and varying the first-order action consistently are equivalent on the regular branch.

We now evaluate the connection response on the tensor-only synchronous path used in Sec.~\ref{sec:tensor_action}. Decompose $\omega_\star=\omega_{\rm LC}(e)+K(e,\chi_\star)$, where $\chi=(V_\mu,A_\mu,Q_{\lambda\mu\nu})$ are the irreducible torsion variables. The Levi--Civita part has its usual metric response and supplies the Einstein--Hilbert tensor operator; the following calculation isolates the algebraic torsion response. On the fully regular branch the unsourced $V_\mu$ and $Q_{\lambda\mu\nu}$ equations are invertible and give zero. For the sourced axial component, take $\bar J_5^\mu=\rho_5\delta^\mu_0$ with $\delta_{\rm TT}\rho_5=\delta_{\rm TT}^{2}\rho_5=0$. Since $g_{00}=-1$, $g_{0i}=0$, and only $g_{ij}$ varies, Eq.~\eqref{eq:app_A_solution} gives
\begin{equation}
\begin{aligned}
\bar J_{5\mu}&=(-\rho_5,0,0,0),\\
A_\mu&=\left(\frac{\kappa\rho_5}{2(1-4\zeta)},0,0,0\right),\\
A^\mu&=\left(-\frac{\kappa\rho_5}{2(1-4\zeta)},0,0,0\right).
\end{aligned}
\label{eq:app_A_TT_components}
\end{equation}
which are independent of $h_{ij}$. Thus
\begin{equation}
\begin{aligned}
\delta_{\rm TT}A_\mu&=-\frac{\kappa}{2(1-4\zeta)}\delta_{\rm TT}\bar J_{5\mu}=0,\\
\delta_{\rm TT}^{2}A_\mu&=-\frac{\kappa}{2(1-4\zeta)}\delta_{\rm TT}^{2}\bar J_{5\mu}=0.
\end{aligned}
\label{eq:app_A_TT_response}
\end{equation}
The full connection is not held fixed: its response is $\delta_{\rm TT}\omega_{\star\mu}{}^{ab}=\delta_{\rm TT}\omega_{{\rm LC}\,\mu}{}^{ab}+\left.\delta_{\rm TT}K_\mu{}^{ab}\right|_{\delta_{\rm TT}\chi=0}$. The tetrad-induced contortion term need not vanish; it is retained by the connection-to-contortion change of variables and the Schur reduction.

To see how this response enters the constrained Hessian, contract Eq.~\eqref{eq:app_schur_proof} with a TT variation. In the axial block, the tensor-only path has $\sqrt{-g}=a^3$, $g^{00}=-1$, and fixed $A_0,J_5^0$, while $A_i=J_5^i=0$. Hence both the direct TT Hessian at fixed $A_\mu$ and the projected mixed block $S_{h_{\rm TT}A_\mu}$ vanish; the axial Schur term $S_{hA}S_{AA}^{-1}S_{Ah}$ therefore vanishes on this subspace. Equivalently, after eliminating $A_\mu$, the contact action is $-3\kappa/[16(1-4\zeta)]\int d^4x\sqrt{-g}\,\mathcal Q_{5,\rm MF}$, with $\mathcal Q_{5,\rm MF}=-\rho_5^2$; both its first and second TT variations vanish. A nonzero TT response of $\mathcal C_5$ would invalidate this cancellation and add the state-dependent term displayed in Eq.~\eqref{eq:connected_tt_response}; its value is not fixed by the one-point current. The only surviving leading two-derivative TT Hessian is consequently the Einstein--Hilbert one under the stated closure, as in Proposition~1. This explicit projection also shows why the full auxiliary connection must be eliminated consistently rather than frozen before varying the tetrad.

\setlength{\bibsep}{0pt}
\renewcommand{\bibfont}{\fontsize{7.5pt}{7.7pt}\selectfont}
\bibliographystyle{spphys}

\bibliography{cas-refs}

\end{document}